# Quantitative Matching of Forensic Evidence Fragments Utilizing 3D Microscopy Analysis of Fracture Surface Replicas

Bishoy Dawood [a], Carlos Llosa-Vite [b], Geoffrey Z. Thompson [b], Barbara K. Lograsso [c], Lauren K. Claytor[d], John Vanderkolk [e], William Meeker [b], Ranjan Maitra [b], and Ashraf Bastawros [a]

[a] Iowa State University, Department of Aerospace Engineering, Ames, Iowa, 50011;

[b] Iowa State University, Department of Statistics, Ames, Iowa, 50011;

[c] Iowa State University, Department of Mechanical Engineering, Ames, Iowa, 50011;

[d] Virginia Department of Forensic Science Richmond, Virginia, 23219;

[e] Retired from Indiana State Police Laboratory, Fort Wayne, Indiana, 46804

## Highlights:

- Silicone casts are widely used by practitioners in the comparative analysis of forensic items.
- 3D topological images of fractured surface pairs and their replicas are analyzed.
- Statistical analysis comparison protocol was applied to provide confidence in the quantitative statistical comparison between fractured items and their replicas.
- Silicon casts accurately replicate fracture surface topological details with a wavelength greater than 20μm.

September 2021

**Abstract**

Fractured surfaces carry unique details that can provide an accurate quantitative comparison to support comparative forensic analysis of those fractured surfaces. In this study, a statistical analysis comparison protocol was applied to a set of 3D topological images of fractured surface pairs and their replicas to provide confidence in the quantitative statistical comparison between fractured items and their replicas. A set of 10 fractured stainless steel samples were fractured from the same metal rod under controlled conditions and were cast using a standard forensic casting technique. Six 3D topological maps with 50% overlap were acquired for each fractured pair. Spectral analyses were utilized to identify the correlation between topological surface features at different length scales of the surface topology. We selected two frequency bands over the critical wavelength (which is greater than two-grain diameters) for statistical comparison. Our statistical model utilized a matrix-variate-*t* distribution that accounts for the image-overlap to model the match and non-match population densities. A decision rule was developed to identify the probability of matched and unmatched pairs of surfaces. The proposed methodology correctly classified the fractured steel surfaces and their replicas with a posterior probability of match exceeding 99.96%. Moreover, the replication technique shows the potential to accurately replicate fracture surface topological details with a wavelength greater than 20μm, which far exceeds the range for comparison of most metallic alloys of 50-200μm. The developed framework establishes the basis of forensic comparison of fractured articles and their replicas while providing a reliable quantitative statistical forensic comparison, utilizing fracture mechanics-based analysis of the fracture surface topology.

# 1 INTRODUCTION

A physical fit or physical match, as described by the American Society of Trace Evidence Examiners (ASTEE), is the alignment between two or more pieces to determine whether they once formed a single object [1, 28]. Matching the physical fractures of different materials such as wood, glass, paper, skin, cables, tapes, and metals has been widely studied [3-17, 21]. This physical matching utilizes the thickness, color, pattern, fracture morphology, irregularities in the fracture, and imperfections across the fracture location [38, 29]. Patterns along the complex jagged trajectory of a macro-crack (large cracks visible to the naked eye) are considered unique and can be utilized to distinguish matching pairs of fractured surfaces by an examiner or by a layperson on a jury [2, 18]. However, reliable examination decisions require experienced forensic experts using comparative microscopy and physical pattern matching. Moreover, the error rate is difficult to quantify in physical matching due to many factors, including fragment material properties, loading and environmental exposure, and forensic scientist judgment and experience [28].

During the last two decades, new innovative 3D surface topological scanning has been developed with the potential to improve physical matching. Different 3D acquisition systems, including 3D laser scan, optical coherence tomography, optical microscopy, laser or light microscopy, stylus scanning instruments, and confocal microscopy, have been utilized for forensic evidence identification applications [38,39,43-45,47-56]. Automated surface acquisition and matching processes utilizing 3D topography data have demonstrated promising improvements in the objectivity of the comparison process [47]. Specifically, confocal microscopy utilizes a pinhole aperture, allowing only light that is reflected from the in-focus plane to where it is captured. Confocal microscopy allows for slices of surfaces to be captured at different depths and then stacked on top of each other to render a 3D image of the surface topology. This 3D image can then be visualized as a two-dimensional profile of the surface roughness, as shown in Fig. 1(c). Zooming out from the profile (looking at longer wavelengths, or spacing between topological events) makes the profile appear smoother, while zooming in on the profile (looking at shorter wavelengths, or spacing between events) results in the 2D profile appearing to be rougher and makes it possible to identify features for forensic comparison. Furthermore, forensic replicas utilizing silicone casts are widely used by practitioners in the comparative analysis of forensic items [21, 55, 32, 33]. However, there is a dearth of in-depth analysis of the ability of a silicon cast to reproduce a useful range of topographical details needed for 3D analysis. Especially, the silicon cast of a fracture surface may not replicate all the small features that represent the short-wavelength topology of the original surface.

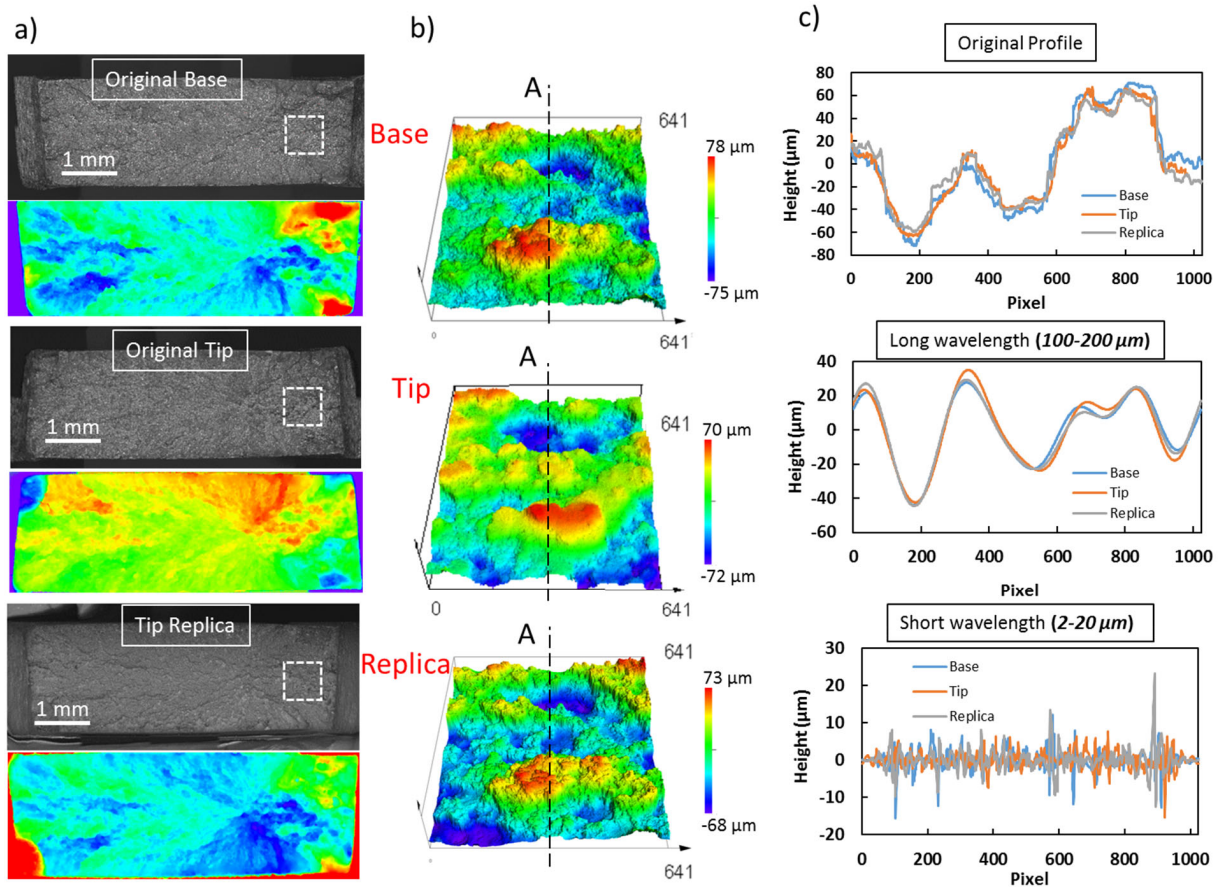

Figure 1: 3D topological analysis of a Base-Tip pair of fractured surfaces and a Tip-cast replica. (a) Optical micrograph and color rendering of the topological fracture surface. Base/Tip and Tip/its cast show mirror symmetry. (b) 3D topological representation of the fracture surface, utilizing a 640 μm Field of View for the square inset area on the optical image. (c) 2D representation of the height comparison along line A-A showing the original measured height (top) and the corresponding long-wavelength (middle) and short-wavelength (bottom), utilizing spectral decomposition by Fourier Transform Analysis. The low-frequency topological details (middle) exhibit patterns that are relevant for statistical comparison while the short-wavelengths topological details (bottom) exhibit no comparable patterns.

In this paper, we address the limits and applicability of surface replica for forensic comparison using our formal quantitative framework to quantify the probability of a match between two specimens in question [41, 46]. We combine fracture mechanics with statistics and machine learning to arrive at a comparison decision. When the 3D spectral analysis of the fractured surface topography is combined with a statistical learning tool, the domain of unique individuality can be easily identified [46]. This tool can provide a quantitative analysis for match probability and the corresponding error rate that is required to be reported [19, 20]. The indefinite microscopic features on the fracture surface topology, as highlighted in Fig. 1(a, b), carry considerable unique details that may be used to support the forensic examiner's decisions with a quantitative forensic comparison. The fractured surfaces show self-affine scaling properties (proportionality of surface roughness with the observation window scale) to quantitatively correlate the material resistance to fracture with the resulting surface roughness, as highlighted in Fig. 2. The limit of the self-affine scale is controlled by the material resistance to fracture, crystal structure, material impurities and loading conditions [12, 22-25]. The surface characteristic becomes unique and non-self-affine at a larger scale, say, $\lambda$ [26, 27], which sets the proper scale for forensic comparison of fractured surfaces. Figure 2 shows the fracture surface topology and the corresponding height-height correlation as a function of the imaging window size. At a larger length scale ($\lambda \sim 60\text{-}70$ μm for the examined SS alloy), the individuality of the fracture surface topology can exist [25, 46], noted by the saturation of the correlation coefficient. This scale is about the size of the fracture process zone ahead of the crack tip, typically extending to 2-3 times the grain size ($d_g \approx 25-35 \mu m$ for the examined SS alloy). Furthermore, Fig. 2 shows also how a well-cast surface replica exhibits similar characteristics to the original surface. On the other hand, a bad surface replica with entrapped air bubbles deviates from the original surface characteristics and exhibits different correlations.

To highlight these characteristic length scales and their impact on the forensic comparison of the fracture surface, Fig. 1(c) illustrates the role of the proper comparison band of the wavelengths (unique character) within the fracture surface topology. The three lines represent the marked line-A at the center of the pair of processed original Base and Tip surface images and the Tip's replicas (see Fig. 1(b)). The pair of images was processed by the mathematical Fast Fourier Transform (FFT) operator. In one set, the low-frequency content (large wavelength, $\lambda > 100 \mu m$) was retained, as highlighted in the middle plot of Fig. 1(c). The three profiles showed a perfect match and asserted the uniqueness of the topology in this range.

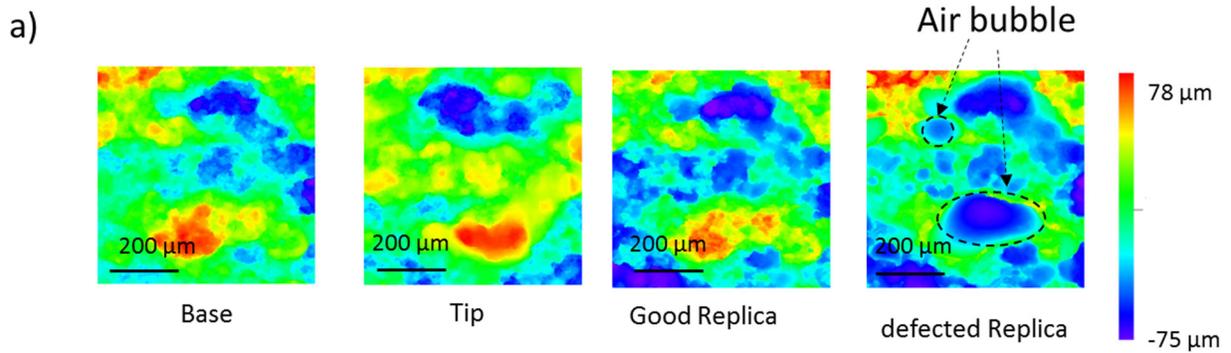

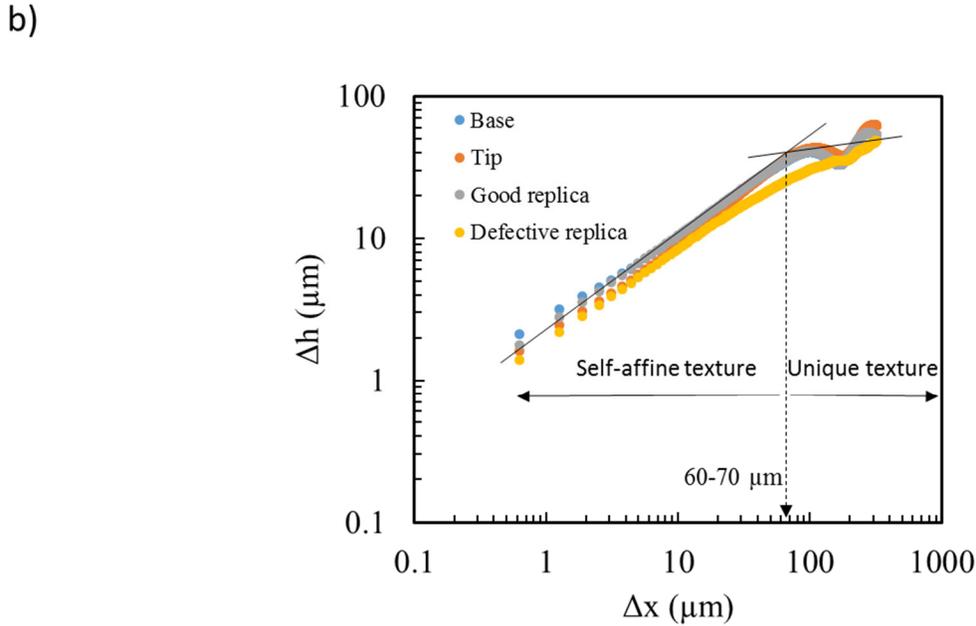

Figure 2: Role of replica quality in identifying unique length scales for fracture comparison. (a) Color map rendering for topological heights, showing a pair of Base-Tip and good vs. defective replicas of the Tip due to large voids. Notice that both the Base and the Tip-replica images are mirror images of the tip image. (b)The corresponding variation of the height-height statistical correlation with the size of the imaging window. The existence of a bubble within the field of view altered the critical wavelength at which the topological details show a transition from self-affine textures to the unique texture that can be used for matching purposes.

In the other set, the high-frequency content was retained (small wavelength, $2\mu m \leq \lambda < 20\mu m$), as highlighted on the lower plot of Fig. 1(c). The line comparison was quite random with no similarity, indicating the self-affine deformation at this range. This trend is similar to that of an optical image obtained by high magnification and a small field of view, where the local fracture mechanism shows similar topological surface features over the fractured surface with indistinguishable character. Accordingly, the utilization of the observation length scale, λ for the unique fractured surface texture provides the proper level of details with the proper microscope magnification.

In this paper, a set of fracture pairs and their replica were examined via 3D imaging profilometry and then analyzed with statistical decision-making tools. We quantify the range of features' resolution that generic silicon-type casting materials can capture to replicate the fracture topology. In particular, we identify the ranges of wavelength and frequencies that it replicates to perform quantitative forensic fracture matches. In optical microscopy, it will set the proper magnification to view the replicated surface and the number of features or events that can be identified for matching purposes. For this purpose, we acquired six overlapping topological images of each fracture surface (Base and Tip) along with the Tip's replicas and obtained correlations for each of the three comparison pairs along two frequency ranges and six overlapping images to obtain a matrix-valued feature that can be used to distinguish between matching and non-matching pairs of images. To classify these matrix features, a quadratic discriminant analysis (QDA) classification algorithm with the matrix-$t$ distribution is trained on a separate set of samples of the same material and imaged by the same operator. This classifier is then used to classify the three surface-pairs: (i) the original Base with Tip, (ii) Tip's casted replicas with original Base, and (iii) Tip's casted replicas with original Tip, resulting in perfect classification in all cases. The details of materials and methods are presented in Section 2, highlighting sample preparation and surface replication, fracture surface imaging protocol, surface spectral analysis, and a summary of the statistical model. The results and findings are summarized in Section 3 and followed by concluding remarks in Section 4.

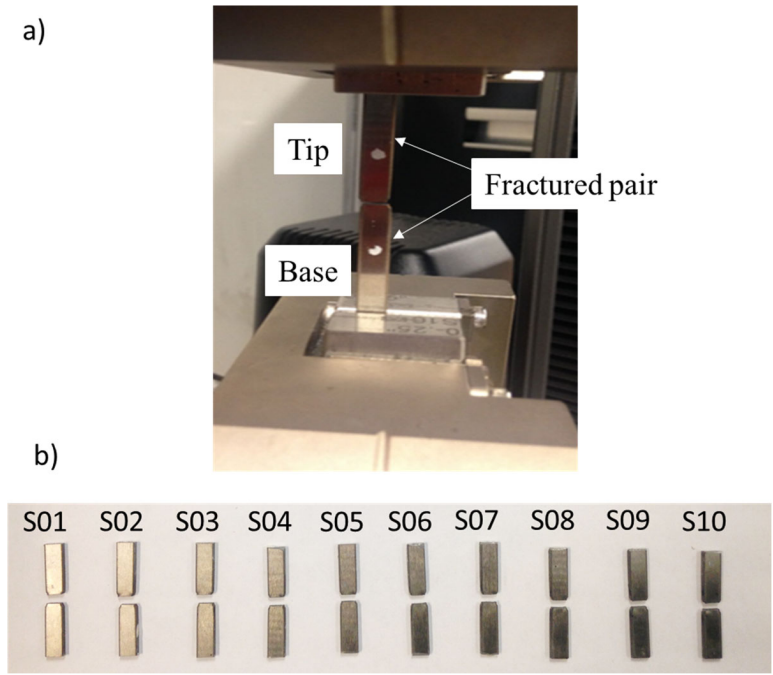

Figure 3: (a) Steel rods' fracture under a controlled tensile loading, (b) A set of 10 fracture-pairs of steel rods utilized in this study.

## 2 MATERIALS AND METHODS

### 2.1 Fracture Samples and Replica Generation

A set of ten rectangular (0.25" wide and 1/16" thick) rods of a common tool steel material (SS-440C) and cut from the same metal sheet to minimize any variability from the manufacturer was used in this study. The steel rods were loaded in an INSTRON 8862 servo-electric computer-controlled testing frame under a controlled extension of 1 mm/min displacement rate till fracture. The fractured pair surfaces were coded as a "Base" and a "Tip," as shown within the loading grips of Fig. 3(a). A set of 10 tool steel rods were fractured, and the fracture-pairs are shown in Fig. 3(b).

A gray silicone type casting material was utilized in generating the fracture surface replicas because it is one of the common replication materials for forensic analysis [55, 33]. The Tip surface was replicated for the entire set. One of the major troublesome issues was the appearance of air bubbles within the image field of view. The air bubbles ranged in size from 70-200 μm, which greatly interfered with the analysis. Figure 2 shows the effect that a bad replica has on biasing the domain of unique surface textures to be used for the comparative analysis. The replica surface with bubbles failed to show unique textures within the field of view FOV and thereby could not be used for the comparative analysis. A glass slide with a uniform layer made of a silicon-based replica material was applied upside down, to the fracture surface, as in Fig. 4(a, b), to replicate the fracture surface. The glass slide was gently eased onto the surface with a 45º inclination to eliminate or minimize the entrapment of air pockets. In order to reduce the viscosity of replica material and improve its flowability and reduce the propensity for bubble entrapment during the replica process, two droplets of acetone were applied to the resin and the hardener mixture. The reduced viscosity mixture greatly enhanced the flowability of the replica paste to capture the fine details of the fracture surface and to provide a bubble-free surface, as shown in the set of images of Fig. 2(a). The replica cast was left on the sample surface for 15 minutes to be fully hardened. The images in Fig.1 (a, b) show a Base-Tip fracture pair and the replica of the Tip, which exactly matches both the Base and the Tip. The Tip for each of the ten fracture pairs was replicated and utilized to form three groups for comparative analysis, namely (i) Base-Tip, (ii) Base-Replicas, and (iii) Tip-Replicas.

**2.2 Fracture Surface imaging**

All 3D fractured surface height topological maps were acquired by a 3D confocal laser microscope (OLYMPUS LEXT-OLS5000). The microscope provides optical magnification of 5-100X with additional digital multiplication of 9X. For the purpose of classification of matching and non-matching surfaces, the observation scales (imaging window) should be controlled by the transition length scale, which should cover at least ten periods of the fracture process zone or about 20-30 grain diameters to avert signal aliasing. The transition length scale is shown in Fig. 2(b), beyond which a saturation in the height-height correlation is seen and indicative of unique surface texture. Accordingly, we utilized 20X objective, which maps the measuring array of 1024 x 1024 pixels to a 640μm FOV with 0.625μm/pixel spatial inter-point resolution.

The topological imaging protocol includes an alignment step and an image-adjustment step. The alignment step ensures that the pair of fractured surfaces to be analyzed are aligned relative to each other without planar misalignment (in-plane tilt of the pair of images), which could significantly deteriorate the surfaces to be compared are viewed simultaneously and tilt adjustment done visually. For 3D topological correlation of the pairs of surface spectra. In optical comparative microscopy, the pair of fractured comparison, such alignment step can be done mathematically [36], adding an extra step of complexity. To streamline the imaging process, a fixture was developed to hold the pair of fracture surfaces parallel and aligned to each other, as shown in Fig. 4(c). The fixture also allows rotational movements around two axes to accommodate non-planar-titled fracture surfaces. The image-adjustment-step entails adjusting the imaging volume to capture the entire topological surface height range of the fractured surface within the field of view. The laser intensity was then adjusted to map the entire surface topology within the dynamic range of the optical sensor without imposing over-saturation and truncation of extreme (high tortuosity) topological details. A standard mathematical out-of-plane tilt is applied to all images to remove global fracture surface tilt. Furthermore, standard mathematical spike noise removal is also applied to remove any measurements that is beyond one standard deviation from the average of the surrounding window of 7-pixel radius. The noisy pixel is replaced with the surrounding window average. It should be noted that the range of wavelength of interests (essential topological features for comparison) are mapped to about 50-100 pixels, rendering the correction window to be less than ¼ of the lowest wavelength and thereby not affecting any topological feature of interest.

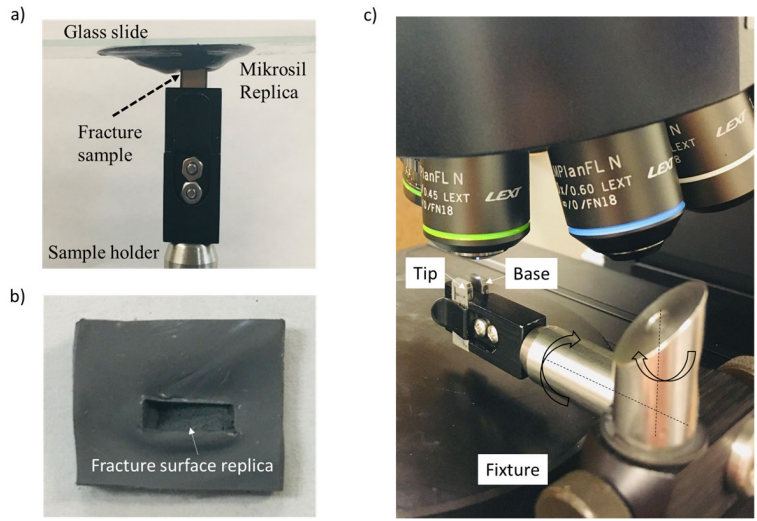

Figure 4: Replication process and imaging of the fracture surface. (a) Vertical replication process of the fracture surface utilizing a gray silicone type casting material, (b) Glass slide containing the replica of the fracture surface, (c) Fracture surface characterization using the laser confocal microscope showing Base and its Tip alignment using the same microscope.

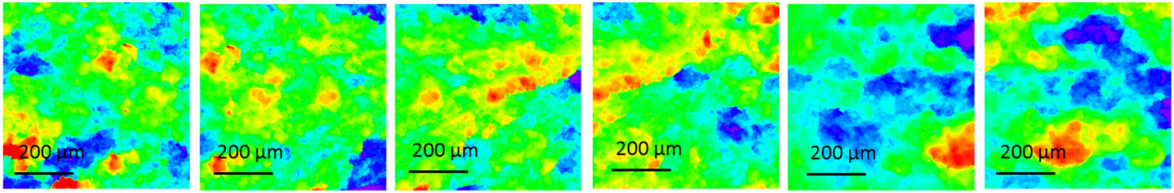
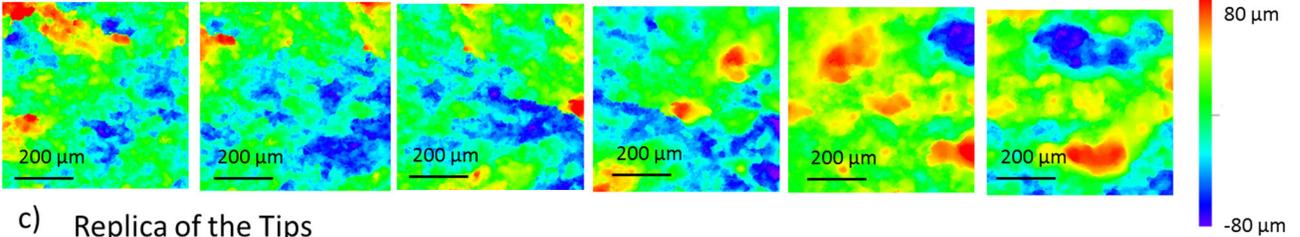
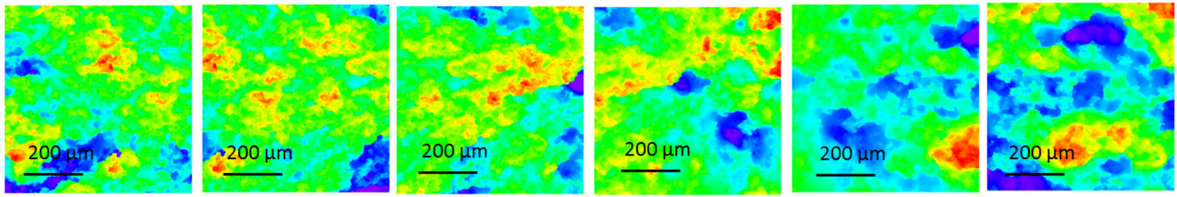

Figure 5: A representative topological set of images for a fracture-pair with acquired 6-images at 50% overlap and 20X objective. (a) Base, (b) Tip, and (c) a replica of the Tip

All images were acquired relative to a reference mark which is the corner of the sample. The starting image, marked by the white box on Fig. 1(a), starting at a 1000μm from the right vertical edge and at the centerline of the sample thickness. For each fracture pair, a set of six successive images with 50% overlap was acquired for both the Tip and the Base. Our previous analysis [46] showed that a set of six images with 50% overlap is required to get classification with very high probability, and to ensure diminishing probability of misclassifications. Figure 5 shows a sequence of six topological images for a Base-Tip fracture pair and the replica of the Tip.

## 2.3 Spectral Analysis and Image Correlation

The FFT operator was applied to each image to generate the corresponding frequency-space representation. A mathematical Hann filter with 10% edge smoothing ratio was applied to the original image to provide a periodic boundary for the image edges, before the FFT operator. The implementation of frequency space analysis provides segmentation of the surface topological frequency ranges for comparison. For instance, the lower frequency bands represent the macro fracture features and the unique river marks.The high frequencies represent the micro mechanism of the fracture process as depicted in the 1-D topological height profiles shown in Fig. 1(c) for line-A showing the original topological details, and the decomposition of the long and short wavelength components. For statistical comparison and decision-making, the statistical correlations between a pair of surfaces' spectra are computed within banded frequencies, with increments in the bands determined by the scale of the image and the microstructure of the material, yielding a similarity measure on each frequency band for corresponding pairs of images.

Image pairs for when the tip and base surfaces were from the same rod are called true matches, while surface pairs from different rods are called true non-matches. Figure 6 shows the correlation distribution for comparison made on pairs of images from matching and non-matching fracture pairs to estimate the distribution for both true matches and true non-matches. The data were derived from 10 base and tip pairs and six images from each surface resulting in 60 matched pairs and 540 non-matched pairs' comparisons. Correlation analysis shows a clear separation between the true matches (red color in Fig. 6) and true non-matches (blue color in Fig. 6) for the 5-10 and 10-20 $mm^{-1}$ frequency bands. The distributions start to be less discriminating and overlap more at higher frequencies. The overlap between matches and non-matches can be further reduced by combining the two most discriminating bands of 5-10 and 10-20 $mm^{-1}$, as seen in Figure 7, which displays the correlations in the Fisher-Z transform (the inverse hyperbolic tangent) scale. According to the contours, the surface-pair with the least overlapping matches and non-matches is replica-tip, which highlights the effectiveness of the surface replicas in capturing the important relevant information for classification.

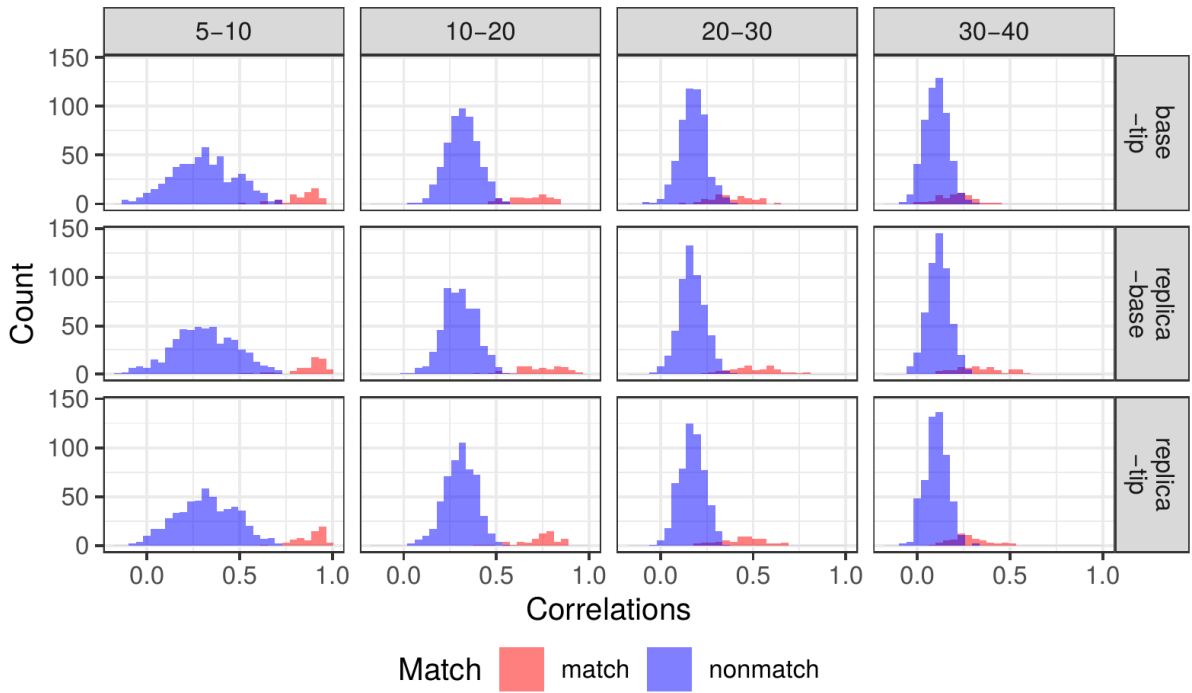

Figure 6: Correlation histograms for true matches and true non-matches for different frequency bands and surface-pair comparison. Lower frequencies are well separated, while higher frequencies start to have overlap.

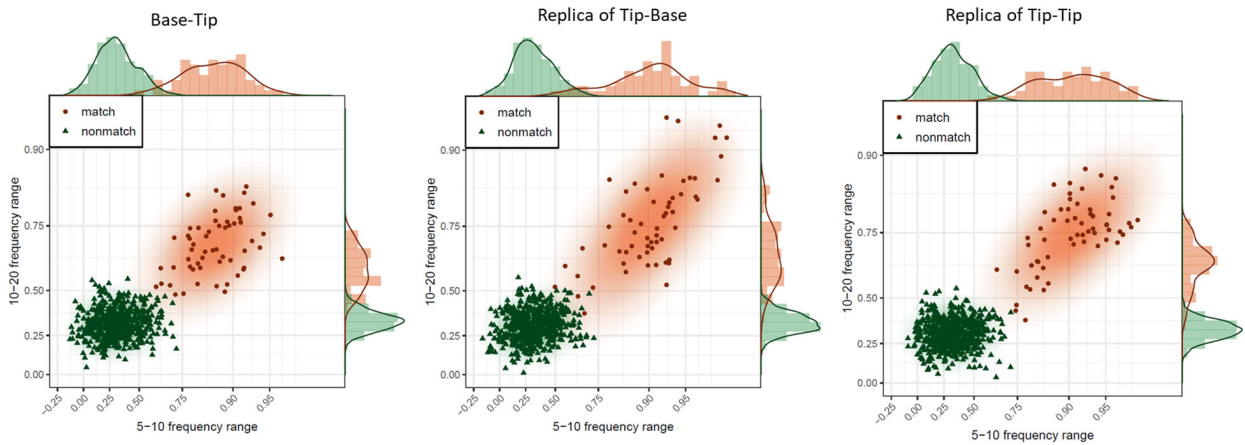

Figure 7: Scatterplots of fracture surface correlations in the Fisher-Z transform scale, and for the three sets of surface-pair comparison. Combining frequency bands increase the overlap between matches and non-matches. In each plot there a total of 60 matched pairs and 540 unmatched pairs.

## 2.4 Model Training and Classification

The correlations that result from comparing each surface pair can be arranged in a 2×6 matrix, where the two rows correspond to the two relevant frequency bands of 5-10 and 10-20 mm$^{-1}$ and the six columns to the six overlapping images taken on the surfaces. To classify a 2×6 matrix X as a match or non-match, we first apply the Fisher-Z transform to each element of X and then use Bayes' theorem to obtain the posterior probability that X is a match as

$$P(X = \text{match}) = \frac{p_1 f_1(X)}{p_1 f_1(X) + (1-p_1) f_2(X)}, \qquad (1)$$

where $f_1$ and $f_2$ are the probability density functions (pdfs) of the true matches and non-matches, respectively, and $p_1$ is the prior probability that X is a match that we set to $p_1 = 0.5$ (the value of $p_1$ only affects the log-odd probabilities by an additive constant that is zero when $p_1 = 0.5$). The posterior probability given in Eq. (1) involves the unknown functions $f_1$ and $f_2$, which we estimate using a training sample of 10 rods of the same material and fractured the same way as the studied steel rods, and that were imaged by the same operator who imaged the steel rods surfaces and their replicas. To estimate $f_1$ and $f_2$ we fitted the matrix-variate-$t$ distribution [40] with 5 degrees of freedom to both the true matches and true non-matches in our training sample using the expectation-maximization algorithm of [41], which generalizes the block-relaxation algorithm of [42]. Our matrix-variate-$t$ model postulates a mean matrix with identical columns an a correlation structure across the six overlapping images that is dictated according to an AR(1) autoregressive process [46].

Having estimated the pdfs $f_1$ and $f_2$, we proceed to estimate the posterior probability that a 2×6 matrix X corresponds to a matching pair of surfaces as per equation (1). In this study, the classification decision was made at the $p = 0.5$ level, meaning that posterior probabilities greater than 0.5 correspond to matching surfaces. The log-odds ratio of the posterior probability in equation (1) can also be converted to a log-likelihood ratio, which can be used as a measure of forensic evidence [28-33].

**2.5 Assessment of Replica Effectiveness in Wavelength Recovery and Topological Mapping**

In the previous section, we correctly classified every image-pair using a QDA classification algorithm based on the matrix-*t* distribution with 5 degrees of freedom. For the purpose of studying the relevant wavelengths recovered from our replicas, we used the matrix-*t* distribution to obtain the mean correlation along the ten frequency bands defined by the thresholds of 3-5-10-20-25-33-50-67-100-133-200 mm$^{-1}$. This corresponds to wavelength bands of 333-200-100-50-30-20-15-10-7.5-5 μm, respectively. For the Tip-Replica pairs, we obtained the correlations from the six overlapping images for all the specified frequency bands to obtain 10 matching pairs and 90 non-matching pairs. Then, we formulated 10x6 matrix features, with the 10 rows corresponding to the frequency bands and the 6 columns to the overlapping images. We fit the matrix-*t* distribution with 5 degrees of freedom, AR(1) correlation structure along the columns to capture the overlapping context of the images, and identical columns across the mean matrix. We note that the underlying fracture process generating the topography was not spatially stationary, meaning that at different positions within the fracture surface, the frequency of the fracture process varied slightly due to the inherent randomness of the microstructure. The FFT operator integrates over the entire 2D space, but because of non-stationarity, there are slight phase differences at different locations of the fracture surface. These local phase differences give rise to destructive interference between frequencies that really are correlated, especially at the high-frequency range. This phenomenon is an inherent limitation of the FFT operator. To study the effect that noise in the Fourier space has on the obtained mean correlations, we applied a 3-point kernel filter to the FFT images before obtaining the matrix features. The kernel is a 3x3 spatial-frequency-sample blur. Then, we fit the previously specified matrix-*t* distribution.

## 3 ANALYSIS OF THE RESULTS

The current framework utilizes a statistical model to produce a likelihood ratio or log-odds ratio of a matching pair or set of pairs; in addition, our model can estimate the probabilities of misclassification. Other probabilistic models such as the likelihood ratio and Congruent Matching Cells are used in many forensic applications such as fingerprint identifications and bullet matching [34-36, 48]. Our focus also is on examining the potential of the replicas to transfer all the topological details required for the analysis at the proper magnifications. The classification model was applied to three sets of pairs of surfaces: the original Bases with Tips, the Bases with the Tip replicas, and the Tips with their replicas. Figure 8 shows the posterior probabilities of match obtained on the three sets of surface pairs in the log-odds scale. Larger posterior log-odds indicate more evidence that a surface pair is a match, whereas lower log-odds indicate more evidence that a surface pair is a non-match. Utilizing the *t*-distribution with 5 degrees of freedom provides great confidence in the discrimination power of the proposed comparison and statistical analysis framework. The model classified the three cases with ten pairs of true matches and the 90 pairs of true non-matches with no false negatives and no false positives. That is, there are a total of 30 pairs of true matches and 270 pairs of true non-matches. The 90 replicas show a high probability of match when compared to the original fractured surfaces. This high accuracy exists for both original Base-Tips, replicas-Bases, replicas-Tips, although the replicas were cast only on the surfaces of the tips. These results demonstrate the ability of the replicas to capture the relevant features that are important for discrimination. Furthermore, for the true match group, the lowest posterior probability was higher than 0.9996, while the highest posterior probability for the true non-match was less than 0.005. The stark difference between the match and non-match probabilities highlights the strength of using the physical basis of fracture mechanics to guide the imaging procedure and construct the statistical discrimination framework.

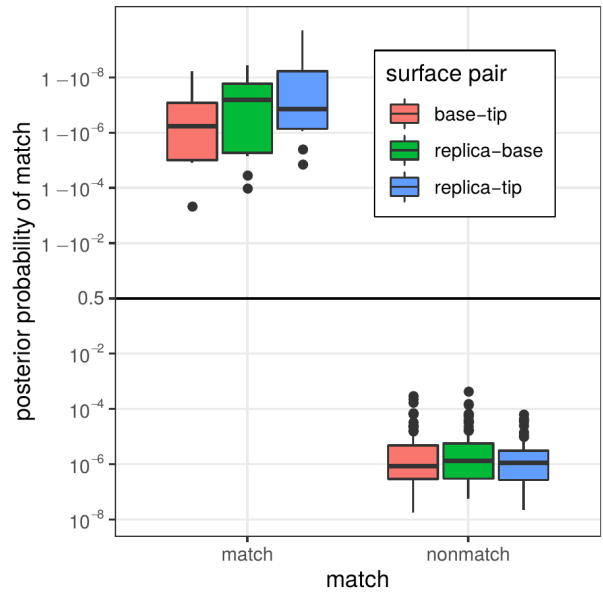

Figure 8: Posterior log-odds of being a match using a model trained on a separate set of images from the same surfaces and tested on the pair of surfaces of base-tip, replica-base, and replica-tip. Higher values indicate stronger evidence of a match.

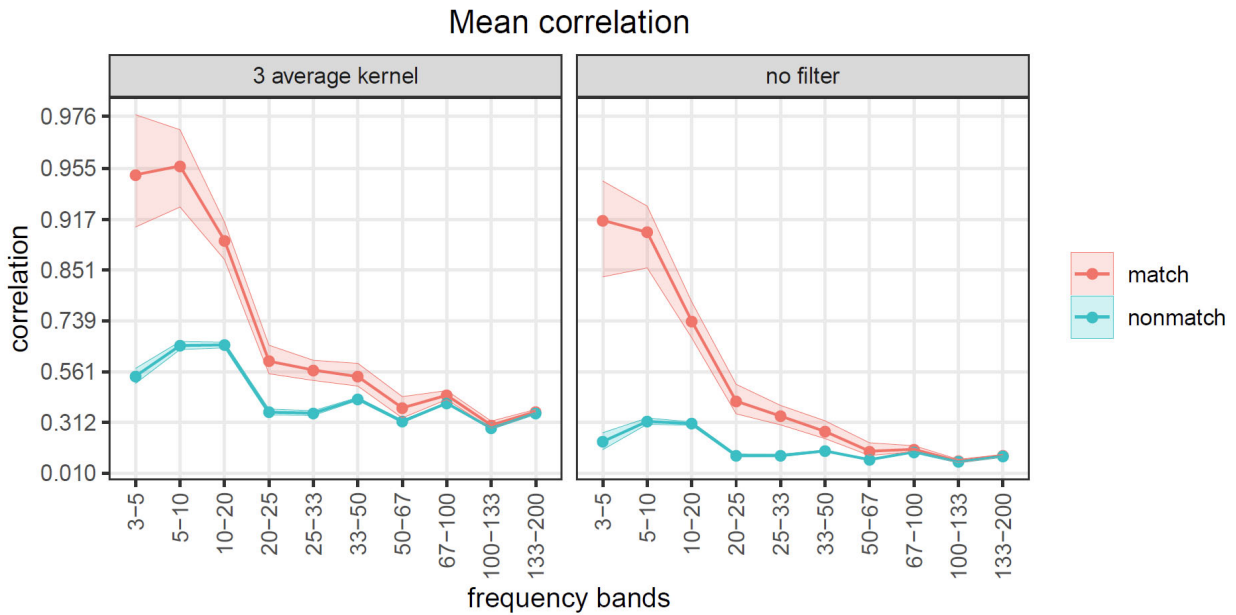

Figure 9: Mean correlations obtained from the comparison between the Tip and its replica along ten frequency bands and to both matches and non-matches, and for both the original and filtered frequency spectra. The matching correlations decrease with increasing frequency bands while still being distinctively different from the non-matches until it reaches frequency bands higher than 100 mm$^{-1}$.

Figure 9 summarizes the results for the replication capacity of the silicone replica technique for all the frequency bands in the range of 3-200 mm$^{-1}$ or the corresponding wavelength of 333-5 μm, respectively. The mean correlations are shown for each of the comparison bands along with 95% bootstrap confidence intervals for the group of match and non-match cases. The results are shown for both the original and the filtered frequency spectra of fractured knives tips and their casted replicas. Figure 9 shows that with increasing frequency bands (i.e., reduction of the wavelength in real space), the match correlations between the Tip and its replica decay in magnitude and spread, indicating loss of differentiation of unique events. Furthermore, the correlations obtained from the filtered FFTs have increased correlations in both the matches and the non-matches. This is equivalent of using a discrimination framework at higher magnification where the fracture surface topology are self-affine and indistinguishable from one surface to another of the same class. However, Fig. 9(a) shows that the match correlation remains around 0.4-0.5 for a high frequency of 100mm$^{-1}$ or a short wavelength of 10μm. This result gives confidence that replicas can still reliably reproduce wavelengths down to the micron ranges. However, such a range needs additional investigation with a smaller FOV and large magnification, employing a master surface with well-defined micron range features. Also, at large wavelengths, that is, at wavelengths greater than about a fifth of the imaging window size, Fig. 9(a) shows a slight drop in correlation at the 3-5mm$^{-1}$ frequency band when compared to the 5-10mm$^{-1}$ frequency band. This is a limitation of the discretization process wherein the resolution per frequency line is 1.56 mm$^{-1}$, which will provide a very limited number of data points (about 7 discretized frequency lines) in the 3-5mm$^{-1}$ band. A larger FOV at the same magnification would be required to refine the frequency band lines and resolve these long-wavelength limitations.

This analysis provides two significant results. *First*, the analysis supports our previously developed classification procedure [46]. It shows how replicas effectively capture surface fractures along wavelength topological details in the range of the two frequency-band analyses of 5-10 and 10-20mm$^{-1}$, equivalently between 200-100 and 100-50μm wavelengths. The topological features at these length scales are unique and helpful for distinguishing between matches and non-matches. *Second*, the analysis shows the ability of the replica to faithfully replicate fracture features with wavelengths all the way to the 25μm range. For forensic comparison, the replicas are well suited for mapping features of 20 μm and larger. It can be assertively stated that the replicas effectively distinguish matches from non-matches in low frequency ranges, and that they stop being distinctively different for frequencies above 100 mm$^{-1}$, where the micro-features of the local fracture processes that are common to both matches and non-matches are compared.

## 4 CONCLUSIONS

We have utilized our developed quantitative statistical comparative analysis framework to examine the potential of the cast replicas in the comparative forensic analysis of topological details of pairs of fractured surfaces. The replica surfaces faithfully reproduce the topological details with wavelength features greater than 20μm. The replicas showed a high probability of match when compared to the original Base-Tip fractured pairs. This result highlights the replicas' ability to capture the relevant features that are important for discrimination. Furthermore, the stark difference between the match and the non-match probabilities highlights the strength of using the physical basis of fracture mechanics to guide the imaging procedure and construct the statistical discrimination framework. The presented framework has a high potential in assisting forensic scientists in providing conclusive decision-making with quantifiable probabilities for a wide range of fractured and broken forensic articles along with their replicas.

All of the classification scores were higher than 99.96%, and were the highest for the Tip-replica comparison pair, demonstrating the potential for using replicas, relative to the original Base-Tip comparison. The underlying correlations, which are strong for the low-frequency bands capturing the macro fracture features, indicate the potential of using replicas to reproduce the relevant features present in forensic fracture evidence.

**ACKNOWLEDGMENTS**

This work was supported by the U.S. Department of Justice under contracts No. 2015-DN-BX-K056 and 2018-R2-CX-0034.

22. Benoit B, Mandelbrot, Passoja DE, and Paullay AJ. Fractal character of fracture surfaces of metals. Nature. 1984; 308:721-722.

23. Ponson L. Crack propagation in disordered materials: how to decipher fracture surfaces. Annales De Physique. 2007; 32(1):1-120.

24. Alava, MJ, Nukala P K V V, and Zapperi S. Statistical models of fracture. Advances in Physics. 2006; 55(3-4):349-476.

25. Bonamy D, and Bouchaud E. Failure of heterogeneous materials: a dynamic phase transition. Physics Reports. 2011; 498(1):1-44.

26. Anderson TL. Fracture Mechanics: Fundamentals and Applications. Academic Press, 2017.

27. Cherepanov, GP, Balankin, AS, and Ivanova VS. Fractal fracture mechanics–a review. Engineering Fracture Mechanics. 1995; 51(6):997-1033.

28. Brooks E, Prusinowski M, Gross S, Trejos T. Forensic physical fits in the trace evidence discipline: A review. Forensic Science International. 2020; 313:110349.

29. Gupta, SR. Matching of fragments, International Criminal Police Review.1970; 198–200.

30. Townshend, DG. Identification of fracture marks, Association of Firearm and Tool Mark Examiners Journal. 1976; 8(2):74-75.

31. Goebel R, Gross KD, Katterwe H, Kammrath W. The comparison scanning electron microscope first experiments in forensic application. Scanning. 1980; 3:193-201.

32. Loci forensics BV. The Netherlands. https://www.lociforensics.nl/forensic-sil/ (accessed June 2021).

33. Moran B. Physical match/tool mark identification involving rubber shoe sole fragments, Association of Firearm and Tool Mark Examiners. 1984; 16(3):126–128.

34. Champod C, Lennard C, Margot P, Stoilovic M. Fingerprints and Other Ridge Skin Impressions. Second Edition. CRC Press. 2016.
22